\documentclass{ws-procs9x6}
\newcommand{\rsun}{R$_\odot$}
\newcommand{\kmps}{kms$^{-1}$}

\usepackage{rotating}
\begin{document}

%\title{High Energy Solar Particle Events and their Associated CMEs}
\title{HIGH ENERGY SOLAR PARTICLE EVENTS AND THEIR ASSOCIATED 
CORONAL MASS EJECTIONS}

\author{P.K. MANOHARAN and G. AGALYA}

\address{Radio Astronomy Centre, National Centre for Radio Astrophysics, \\
Tata Institute of Fundamental Research, Udhagamandalam (Ooty), India.}

%\maketitle

\begin{abstract}

Intense solar energetic particle (SEP) events data, associated with
ground level enhancements (GLEs), occurred during 1989 to 2006
have been obtained from the spectrometers on board GOES spacecraft 
in the energy range 10--100 MeV. The interplanetary effects of
these events and their associated coronal mass ejections (CMEs)
have been provided by the LASCO/SOHO coronagraph images in the 
field of view of 2--30 {\rsun} and the interplanetary scintillation 
images from the Ooty Radio Telescope in the heliocentric 
distance range of $\sim$40--250 R$_\odot$. The comparison between 
the radial evolution of the CME and its associated particle spectrum
shows that the spectrum is soft at the onset of the particle event.
A flat spectrum is observed at the peak of the particle event and
the spectrum becomes steeper as the CME moves farther out into the 
inner heliosphere. However, the magnitude of change in spectral slopes
differs from one CME to the other, suggesting the difference in energy 
available within the CME to drive the shock. The spectral index evolution as
a function of initial speed of the CME at different parts of the 
particle profile has also been compared. The result shows that the 
change in particle flux with time is rather quick for the 
high-energy portion of the spectrum than that of the low-energy part,
which makes the steepening of the energy spectrum with time/distance 
from the Sun. It indicates that the acceleration of particles by a
CME-driven shock may be efficient at low energies ($\leq$30 MeV) and
the efficiency of the shock decreases gradually towards the high-energy 
side of the spectrum.
\end{abstract}


\section{Ground Level Enhancements and SEP Events}

The physical mechanisms involved in the production and acceleration
of solar energetic particle (SEP) events are rather complex and not
yet fully understood. In general, SEP events are associated with
explosive phenomena taking place on the Sun, such as flares and coronal
mass ejections (CMEs).
There are two types of SEP events, impulsive and gradual events
\cite{kall03}{}. Impulsive events have their origin at
the rapid and short-lived (duration $\leq$30 min) flare events,
which are dominated by electrons, $^3$He, and heavy ions. The energy 
release processes associated with the flare play a key role in the 
impulsive acceleration of particles. In the case of gradual events, 
they have association with CME/flare events, which are rich in protons, 
and the CME-driven shocks accelerate particles in the interplanetary 
space\cite{ream97}{}. There are several excellent articles and reviews 
on solar particle events \cite{ream99,ream10,miro08,shea90}{}.

\begin{figure}[t]
\centering
\psfig{file=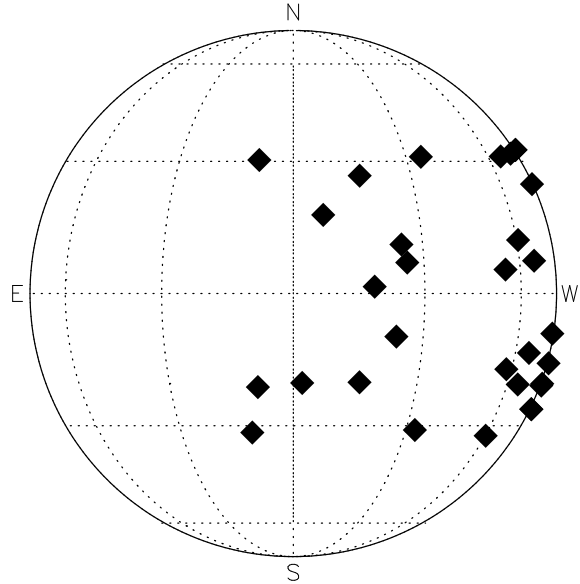,width=7.7cm,angle=-90.0}
\caption{Location of SEP events associated with GLEs observed 
between 1989 and 2006. The position of origin of all these events 
located to the western side of the Sun essentially suggests the 
magnetic connectivity between the source location of particle 
event and the observing point near the Earth.}
\end{figure}

A large and energetic flare/CME associated with an intense
gradual SEP event is important because it can give rise to cosmic
ray ground level enhancements (GLEs). For example, GLEs involve the
hard spectrum of protons (at energies $>$1 GeV) and can be detected
by neutron monitor on the Earth. In this study, we concentrate on 
the evolution of intense particle events in the Sun-Earth distance.
For example, the GLE can be used as an indicator of the associated 
intense particle event. We have selected 31 SEP events associated 
with GLEs in the period 1989 to 2006. The GLE has been only used 
as an indicator for intense particle event. The main focus is however 
on the comparison of temporal variation of energy spectrum of SEP events 
with the propagation properties of CMEs in the near-Sun region as well 
as in the Sun-Earth interplanetary space.

\section{Event Selection and Analysis}

In this study, we consider 31 SEP events associated with ground 
level enhancements (GLEs) occurred between 1989 and 2006. This
period covers the second half of solar cycle 22 and the full extend 
of cycle 23. The solar proton measurements in the energy range of 
10--100 MeV have been obtained from Geostationary Operational 
Environmental (GOES) spacecraft. The near-Sun images of CMEs 
associated with these SEP events, observed after the year 1997, 
have been obtained from the LASCO/SOHO C2 and C3 coronagraphs 
\cite{brue95}{}, which cover the field of view of 2--30 {\rsun} 
(1 solar radius, {\rsun} = 6.96 $\times$ 10$^5$ km).  Whenever 
available, the interplanetary scintillation measurements at Ooty 
on a grid of large number of radio sources provide images of CME 
associated disturbances in the interplanetary space at different 
heliocentric distances until the arrival of the CME at the orbit 
of the Earth (i.e., up to $\sim$250 {\rsun})\cite{mano06}{}. 
The scintillation images in particular are extremely useful in 
understanding the radial evolution of CME in the inner 
heliosphere. 

Table 1 shows the list of 31 SEP events under consideration. It includes 
conventional GLE number ({\it http://neutronm.bartol.udel.edu/}), 
classification of associated flare event, its location on the Sun, 
start of X-ray flux, CME onset time 
at the LASCO C2 coronagraph, type of the CME (full halo or partial halo), 
speed of the CME in the LASCO field of view, shock arrival at 1 AU, and 
start of the particle event as recorded by GOES spectrometer. As shown in 
the table, most of them have association with intense flares (except 3 
events occurred at longitude $>$90$^\circ$ west and they can not be 
accurately classified) and wide CMEs (width $>$180$^\circ$).  Figure 1 
displays the flare location of events listed in Table 1. The remarkable crowding of 
SEP events to the western hemisphere of the Sun (at $>$10$^\circ$  east 
longitude) is consistent with the magnetic connectivity between the 
particle acceleration site and the Earth, i.e., along the Archimedian 
spiral \cite{ream96}{}.

\begin{figure}[t]
\centering
\psfig{file=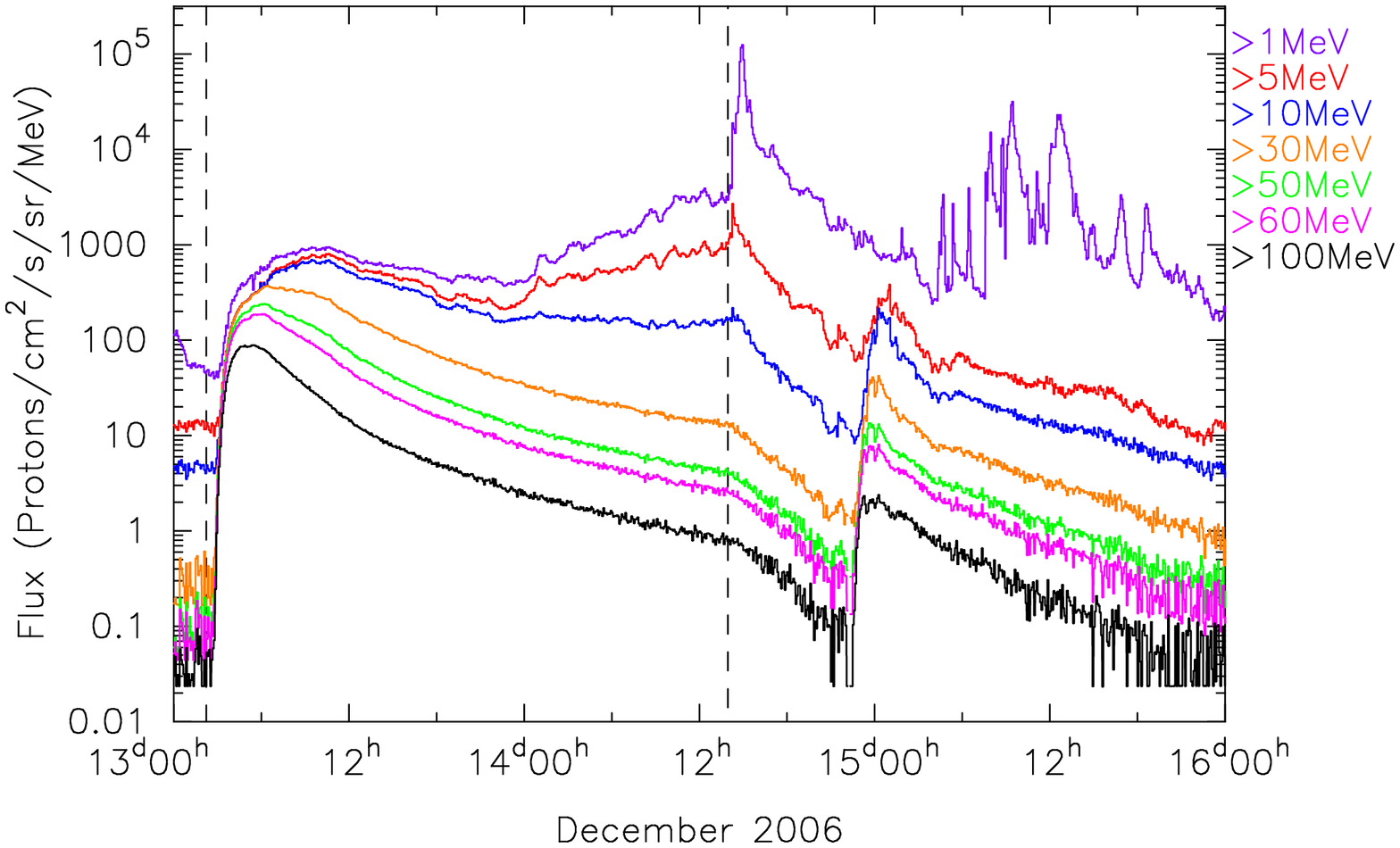,height=11.0cm,angle=-90}
\caption{
GOES-11 proton flux measurements between 13 and 16 December 2006.
The vertical line at 02:14 UT on December 13 indicates the 
time of X-ray onset of the flare event. Another vertical  
line identifies the arrival of interplanetary shock at 1 AU 
on December 14, 2006, at 13:57 UT. The particle energy increases 
from top to bottom curves (refer to right-hand side legend). 
}
\end{figure}

%\begin{figure}
%\centering
%\psfig{file=fig_3.eps,width=7.0cm}
%\caption{The CME onset at the LASCO C2 coronagraph observed on
%December 13, 2006, at 02:54 UT. The CME originated on the Sun
%at S05W23. Its plane of sky speed was $\sim$1775 {\kmps}. The 
%above image is the running difference image, i.e., from the
%image taken at 02:54 UT, the previous background image has been
%subtracted.}
%\end{figure}

\begin{sidewaystable}
\tbl{List of 31 SEP Events}
{\tablefont
\begin{tabular}{@{}l c c c c c c c c c c c@{}}
%\multicolumn{12}{c}{Table 1. List of 31 SEP Events}\\%\hline
\toprule
 && Date & \multicolumn{3}{c}{Flare Data} & \multicolumn{3}{c}{CME Data}
& \multicolumn{2}{c}{Shock at 1 AU} & Particle Event \\ \\
\cline{4-6} \cline{7-9} \cline{10-11}\\
No.& GLE & & Class & Location & Start & C2 Onset &H/PH$^*$& Speed &
Date & Time & Start Time\\
&No.& & & & hh:mm & hh:mm& &kms$^{-1}$ & & hh:mm & hh:mm \\
\hline
1 &  40& 25Jul89 & X2.6/2N &   N25W84 & 08:39 &      &   &     &   26Jul(?) & 13:00(?)&  09:00  \\
2 &  41& 16Aug89 & X20/2N  &   S15W86?& 01:08 &      &   &     &   17Aug    & 15:41      \\
3 &  42& 29Sep89 & X9.8/1B &   S26W90 & 10:47 &      &   &     &   30Sep    & 17:16  &   12:05  \\
4 &  43& 19Oct89 & X13/4B  &   S27E10 & 12:29 &      &   &     &   20Oct    & 09:16  & 13:05  \\
5 &  44& 22Oct89 & X2.9/2B &   S27W31 & 17:08 &      &   &     &  \\
6 &  45& 24Oct89 & X5.7/3B &   S30W57 & 17:36 &      &   &     &   26Oct    & 14:27       \\
7 &  46& 15Nov89 & X3.2/3B &   N11W26 & 06:38 &	     &   &     &   17Nov    & 09:25  &  07:35  \\
8 &  47& 21May90 & X5.5/2B &   N35W36 & 22:12 &	     &   &     &   25May    & 05:10  &  23:55  \\
9 &  48& 24May90 & X9.3/1B &   N33W78 & 20:46 &      &   &     &   26May    & 20:37  &  21:25  \\
10 & 49& 26May90 & X1.4    &    W104? & 20:45 &      &   &     &  \\
11 & 50& 28May90 & C 1.4   &    W130? & 04:28 &	     &   &     &   30May    & 09:04  &  07:15  \\
12 & 51& 11Jun91 & X12/3B  &   N31W17 & 02:09 &	     &   &     &   12Jun    & 10:12  &    \\
13 & 52& 15Jun91 & X12/3B  &   N33W69 & 06:33 &	     &   &     &   17Jun    & 10:19  \\
14 & 53& 25Jun92 & X3.9/2B &   N09W67 & 19:47 &	     &   &     &   27Jun    & 20:35  &  20:45  \\
15 & 54& 02Nov92 & X9      &          & 02:31 &	     &   &     &   04Nov    & 13:12   \\
16 & 55& 06Nov97 & X9.4/2B &   S18W63 & 11:49 & 12:10&H  &1556 &   09Nov    & 10:00  &  13:05  \\
17 & 56& 02May98 & X1.1/3B &   S15W15 & 13:31 & 14:06&H  &938  &   04May    & 02:03  &  14:20  \\
18 & 57& 06May98 & X2.7/1N &   S11W65 & 07:58 & 08:29&PH &1099 &   08May    & 09:20  &  08:45  \\
19 & 58& 24Aug98 & X1/3B   &   N35E09 & 21:50 &      &   &     &   26Aug    & 06:36  &  23:55  \\
20 & 59& 14Jul00 & X5.7/3B &   N22W07 & 10:03 & 10:54&H  &1674 &   15Jul    & 14:35  &  10:45  \\
21 & 60& 15Apr01 & X14.4/2B&   S20W85 & 13:19 & 14:06&PH &1199 &   18Apr    & 00:51  &  14:10  \\
22 & 61& 18Apr01 & C2.2    &   S20WL  & 02:11 & 02:30&H  &2465 &   21Apr    & 15:30  &  03:15  \\
23 & 62& 04Nov01 & X1/3B   &   N06W18 & 16:03 & 16:35&H  &1810 &   06Nov    & 01:45  &  17:05  \\
24 & 63& 26Dec01 & M7/1B   &   N08W54 & 04:32 & 05:30&PH &1446 &   29Dec    & 06:20  &  06:05  \\
25 & 64& 24Aug02 & X3/1F   &   S08W81 & 00:49 & 01:27&H  &1913 &   26Aug    & 11:40  &  01:40  \\
26 & 65& 28Oct03 & X17.2/4B&   S16E08 & 09:51 & 11:30&H  &2459 &   29Oct    & 05:58  &    \\
27 & 66& 29Oct03 & X10/ 2B &   S15W02 & 20:37 & 20:54&H  &2029 &   30Oct    & 16:20  &    \\
28 & 67& 02Nov03 & X8.3/2B &   S14W56 & 17:03 & 17:30&H  &2598 &   04Nov    & 05:53  &    \\
29 & 68& 17Jan05 & X3.8    &   N15W25 & 06:59 & 09:54&H  &2547 &  &  \\
30 & 69& 20Jan05 & X7.1/2B &   N14W61 & 06:36 & 06:54&H  &2400 &   21Jan    & 16:48  &    \\
31 & 70& 13Dec06 & X3.4/4B &   S05W23 & 02:14 & 02:54&H  &1774 &   14Dec    & 13:57  &  03:10  \\
\Hline
\end{tabular}}
%\multicolumn{8}{l}{* H - Halo CME; PH - partial halo CME.}\\
\begin{tabnote}
$^{\text *}$ H - Halo CME; PH - partial halo CME.\\
\end{tabnote}
\end{sidewaystable}

\section{Particle Data and CME Speed}

Figure 2 shows an example of gradual SEP event, observed in 
association with a fast halo CME and an intense flare (X3.4/4B) 
that occurred on December 13, 2006 at S05W23 on the Sun. The
X-ray onset of the flare at 02:14 UT on December 13 and the
1-AU arrival time of the CME-associated shock at 13:57 UT on
December 14 are indicated by respective vertical lines on the
plot. The low-energy protons (E $<$10 MeV) show gradual 
increase after about 0 UT on December 14, suggesting an increase 
in the efficiency of the shock-driven acceleration. Additionally, 
on the arrival of the interplanetary shock at the near-Earth space, 
low-energy protons show a sharp increase in flux, which is due 
to the increase in the population of energetic storm particles. 
However, after the passage of the shock all the energy channels 
show systematic decrease in proton flux. 

In association with this intense flare 
event, LASCO/SOHO C2 coronagraph recorded the onset of a full halo 
CME at 02:54 UT on December 13. The second order polynomial fit to 
the height-time plot obtained from C2 and C3 images indicates a 
decrease in the sky-plane speed from 2000 to 1700 {\kmps} in the 
distance range of 2--20 {\rsun} \cite{mano10}{}.

\begin{figure}[t]
\centering
\psfig{file=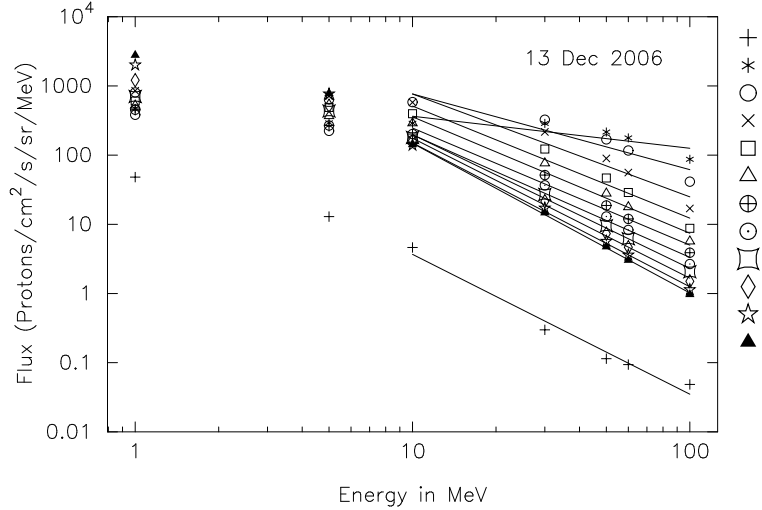,height=10.0cm,angle=-90}
\caption{The energy spectrum of the SEP event associated with the flare 
event on December 13, 2006. Each line in the plot is the least square 
fit to the proton measurements between 10 and 100 MeV. The symbols 
displayed at the right-hand side of the plot, from top to bottom 
respectively, represent 3-hour interval fits from the flare onset time 
to shock arrival at the Earth. It is evident that the overall particle 
flux decreases in all energy bands also the spectrum steepens as the 
CME moves away from the Sun.}
\end{figure}

\section{SEP Energy Spectrum}

As shown in Table 1, each SEP event was accompanied by a fast and wide CME.
At the leading edge of the CME, a shock wave was observed and part
of it passed through the Earth-orbiting satellite as an interplanetary
shock. The arrival time of the shock at 1 AU is included in Table 1.
For each event, based on the GOES spectrometer data, we have 
computed the energy spectrum of protons (E$^{\gamma}$), in the energy 
range 10--100 MeV, nearly from the onset time of the particle event 
to the shock arrival time at the near-Earth space, approximately at every 
3-hour interval. We need to consider the dispersion effects on the arrival 
time of particles from the acceleration site (i.e., at the shock front) to
1 AU. The travel time of the particle depends on its propagation 
path, the pitch angle and speed. For a typical background solar wind 
speed of $\sim$400 {\kmps}, the spiral length corresponds to $\sim$1.2 AU in 
the Sun-Earth distance. It is known that high-energy particles 
($>$10 MeV) experience relatively less pitch angle scatter than 
low-energy particles along the interplanetary magnetic field and 
therefore propagate more directly to 1 AU\cite{posn03}{}.
For all the events (except January 17, 2005), we have considered the 
spectral shape at each 3-hour interval on or after the peak of the 
particle profile. For the event on January 17, 2005, the peak of the particle 
profile is observed after $\sim$9 hours of the flare onset.

\begin{figure}[t]
\centering
\psfig{file=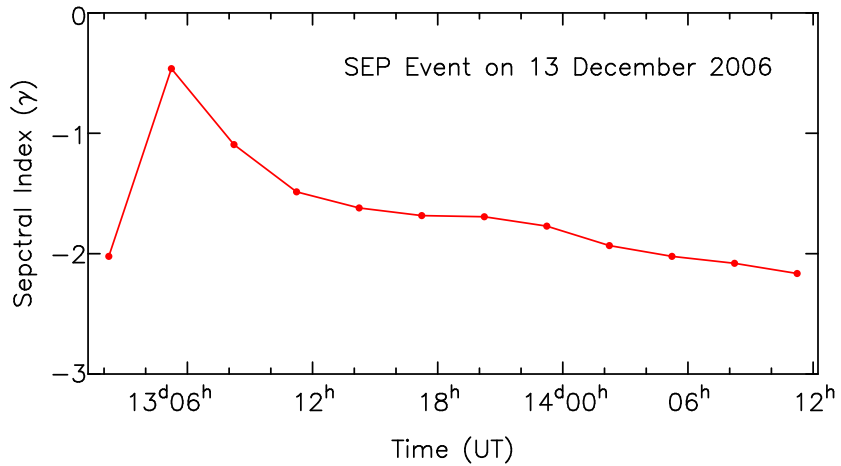,height=11.0cm,angle=-90}
\caption{Spectral index as a function of time for the SEP event on 
December 13, 2006. The spectral index has been obtained in the energy
range 10--100 MeV. The spectrum is flat ($\gamma$ = -0.5) at the peak 
of the particle profile and spectrum steepens with time or as the CME
propagates away from the Sun. }
\end{figure}

In order to study the spectral changes at every 3-hour interval over
an energy range of 10--100 MeV, a least-square fit has been made to 
the {\it log(flux)-log(energy)} plot to get the power-law form of the
spectrum ($\sim$E$^{\gamma}$). However, for some of the intervals, a 
simple power-law (i.e., single straight-line fit) was not adequate. 
Since the primary aim is to study the change in the overall spectral 
shape as a function of time, an average power-law fit has been made to 
the spectrum in the range 10--100 MeV.  Figure 3 shows the spectral fittings in 
the energy range of 10 to 100 MeV at consecutive 3-hour intervals for the 
SEP event observed on December 13 and 14, 2006. At the initial phase of the 
SEP event, i.e., just after the particle onset, the spectrum looks soft and 
the particle flux at high-energy portion increases with time.

\subsection{Radial Evolution of SEP Spectrum}

In Figure 4, spectral indexes observed at consecutive 3-hour intervals 
are plotted for the SEP event associated with the flare event on 
December 13, 2006. This plot includes spectral index nearly from the 
particle onset time to the shock arrival at the Earth. The spectrum 
attains the maximum flatness at the peak of the particle profile, 
spectral index, $\gamma$ $\approx$ -0.5, and the spectral index 
decreases as the CME propagates away from the Sun. 

\begin{figure}
\centering
\psfig{file=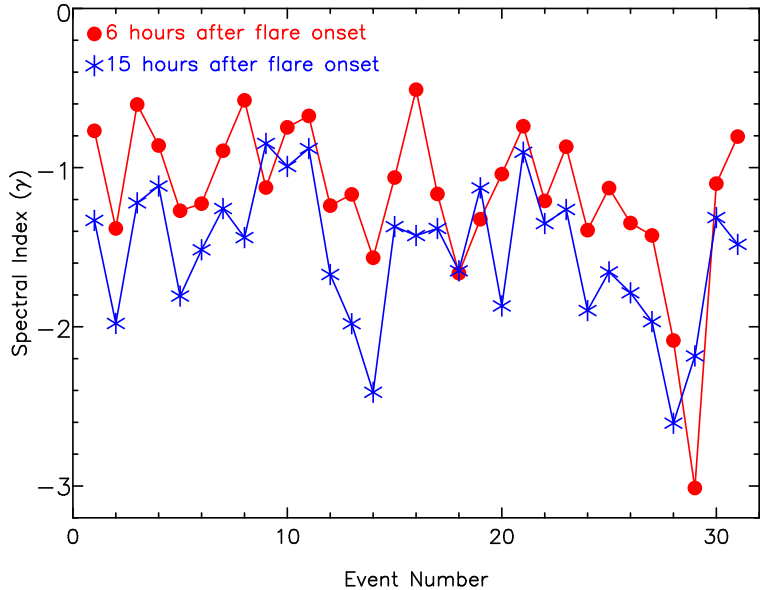,height=10.0cm,angle=-90}
\caption{Spectral index as a function of event number. The x-axis gives 
the event numbers as listed in Table 1. The filled-circle and star 
symbol represent, respectively, the spectral indexes at 6 and 15 hours after 
the flare onset. It is evident from these plots that most of the
spectra (28 out of 31 events) are steeper at large distance from the 
Sun.}
\end{figure}

Another interesting point is that the particle flux at all energy 
bands decreases with time. It is likely linked to the shock ahead of the 
CME event. As the CME propagates, the shock strength (as well as the 
compression ratio in the sheath region) decreases with time/heliocentric distance. 
However, the difference in reduction of particle at the low-energy side of the 
energy spectrum (i.e., $\sim$10 MeV) is less between the peak of the 
profile and the shock arrival time (i.e., typically about an order of magnitude 
reduction is observed). Whereas at the high-energy portion of the 
spectrum (i.e., $\sim$100 MeV), the decrease in particle flux is more 
than or $\sim$2 orders of magnitude. This suggests that only soft protons 
are produced at the front of the CME-driven shock as the CME propagates 
to large distance.  The systematic changes in the particle flux and 
spectral index suggest a decrease in the efficiency of acceleration by 
the shock driven at the front of the CME. The above observed trend is 
compared with the speed evolution of the CME in the Sun-Earth distance. 

\section{Spectral Evolution with Time/Heliocentric Distance}

Figure 5 shows the plots of spectral indexes of 31 SEP events at
6 and 15 hours after the flare onset time. It is evident that out 
of 31 SEP events, 28 of them show steeping of the energy spectrum 
with time as the CME propagates to larger distance from the Sun.
However, the relative steepening differs from one CME to the other,
suggesting that the CME-driven shock or energy available for the
acceleration of particle is CME dependent, which can be mostly 
related to the magnetic energy possessed by the CME as well as
the expansion rate of the CME.

\begin{figure}[t]
\centering
\psfig{file=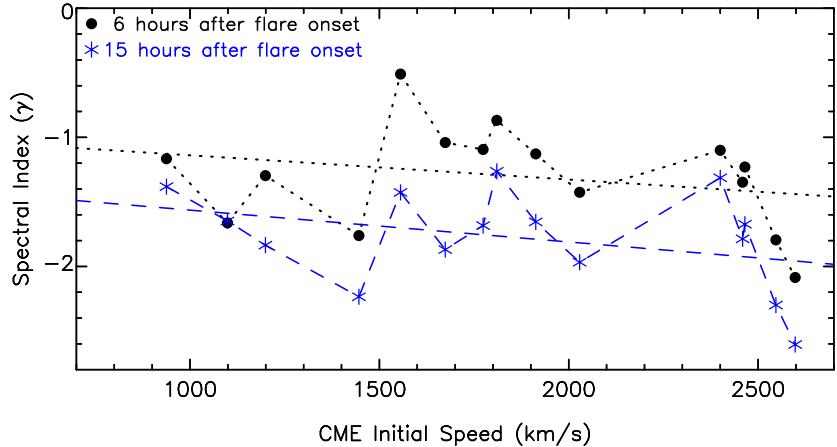,height=11.0cm,angle=-90}
\caption{Spectral indexes at 6 hour and 15 hour after the flare
event are plotted against initial speed of the CME. The initial
speeds have been obtained from LASCO C2 and C3 white-light images
and cover a range of 900 to 2600 {\kmps}. As shown in Figure 5, 
the energy spectrum is flat at near the Sun (6 hours after the 
flare onset) and becomes steep as the CME propagates to large 
distance (15 hours after the flare onset). The least square straight
line fits to 6-hour and 15-hour indexes are shown, respectively, by
dotted and dashed lines.
}
\end{figure}

The other 3 cases showing flattening at large distances are on 24 
May 1990 (event \#9), 24 August 1998 (event \#19), and 17 January
2005 (event \#29). As it is clear from Figure 5, events \#9 and 
\#19 go through only a marginal flattening, where as the event 
\#29 on January 17, 2005, shows heavy flattening with distance from the
Sun. It is to be noted that in this SEP event, the particle flux 
peaked at about 9 hours after the flare peak.  Therefore the
spectral index measured at 6 hours after the flare event is associated
with the growth phase of the particle profile. It is also be noted
that the CME on January 17, 2005, shows likely interaction with a
preceding CME, which originated nearly from the same location. It is
inferred that the particle profile of this event has been heavily 
influenced by the effects of interaction \cite{gopa02}{}.

\section{CME Speed and Spectral Index Changes}

In this study, we have considered CMEs observed between 1989 and 
2006. However, the speed measurements are available for CME events 
observed after the year 1997 (i.e., after the advent of LASCO/SOHO
mission). The LASCO C2 and C3 white-light images provide the initial 
speed of the CME at distances $\le$30 {\rsun}. In 
Figure 6, we plot the spectral index as a function of initial speed 
of the CME. The spectral indexes after 6 hours of the flare event are 
shown by filled-circle symbols and star symbols represent spectral shape
after 15 hours of the flare onset. As indicated by Figure 5, the 
energy spectrum is steeper at the large distance from the Sun than 
that of its starting phase close to the Sun and the same trend is 
observed for the range of initial speeds between 900 and 2600 {\kmps}. 
The straight lines shown in the plot are least square fits to the
indexes at 6-hour (dotted line) and 15-hour (dashed line), respectively. 
These fits show marginal steepening with speed. However, at speeds 
above 2500 {\kmps}, two events show steeper spectra, which require
more investigations.

\begin{figure}[t]
\centering
\psfig{file=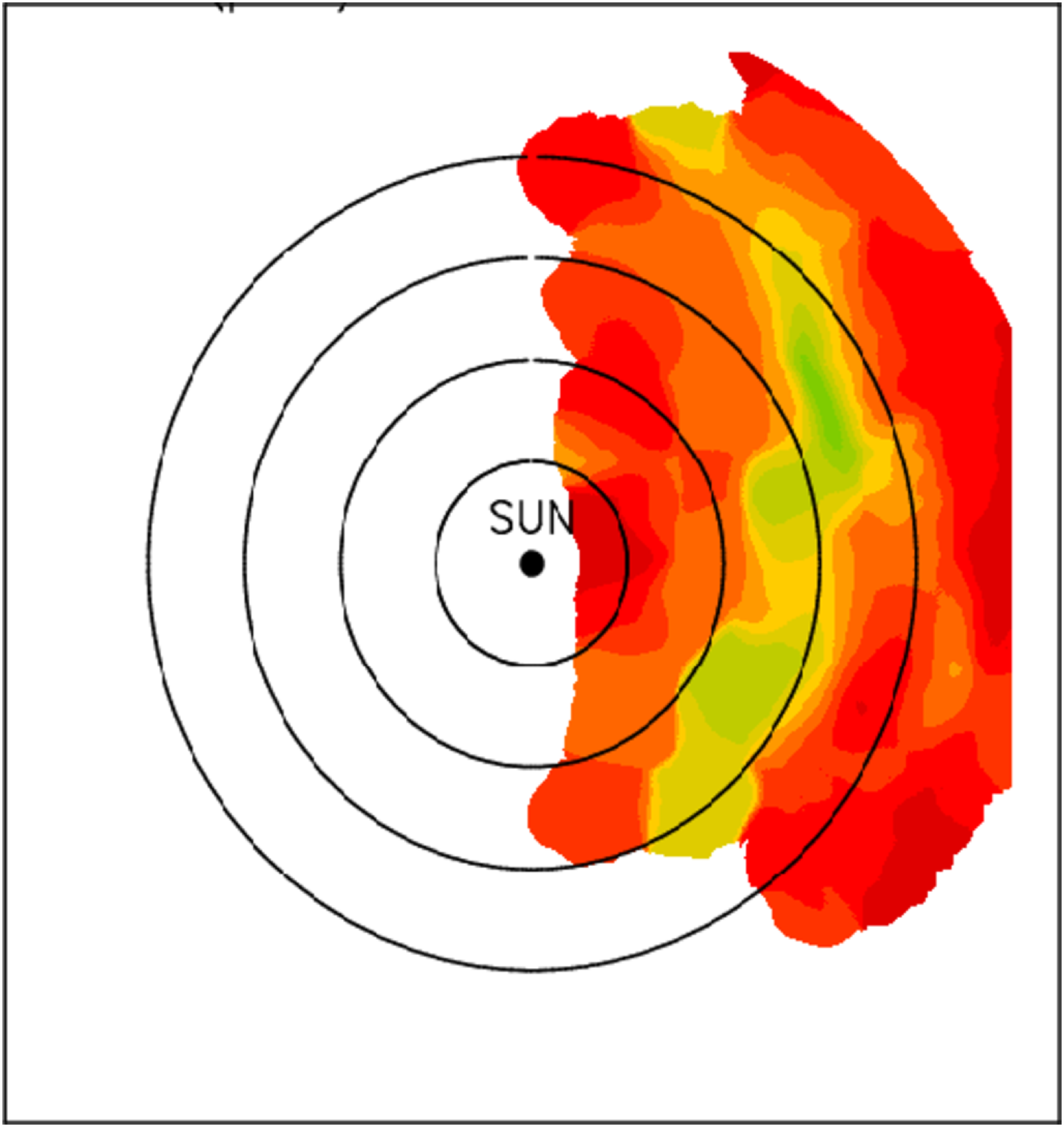,height=4.84cm,width=4.81cm}\hspace{0.95cm}
\psfig{file=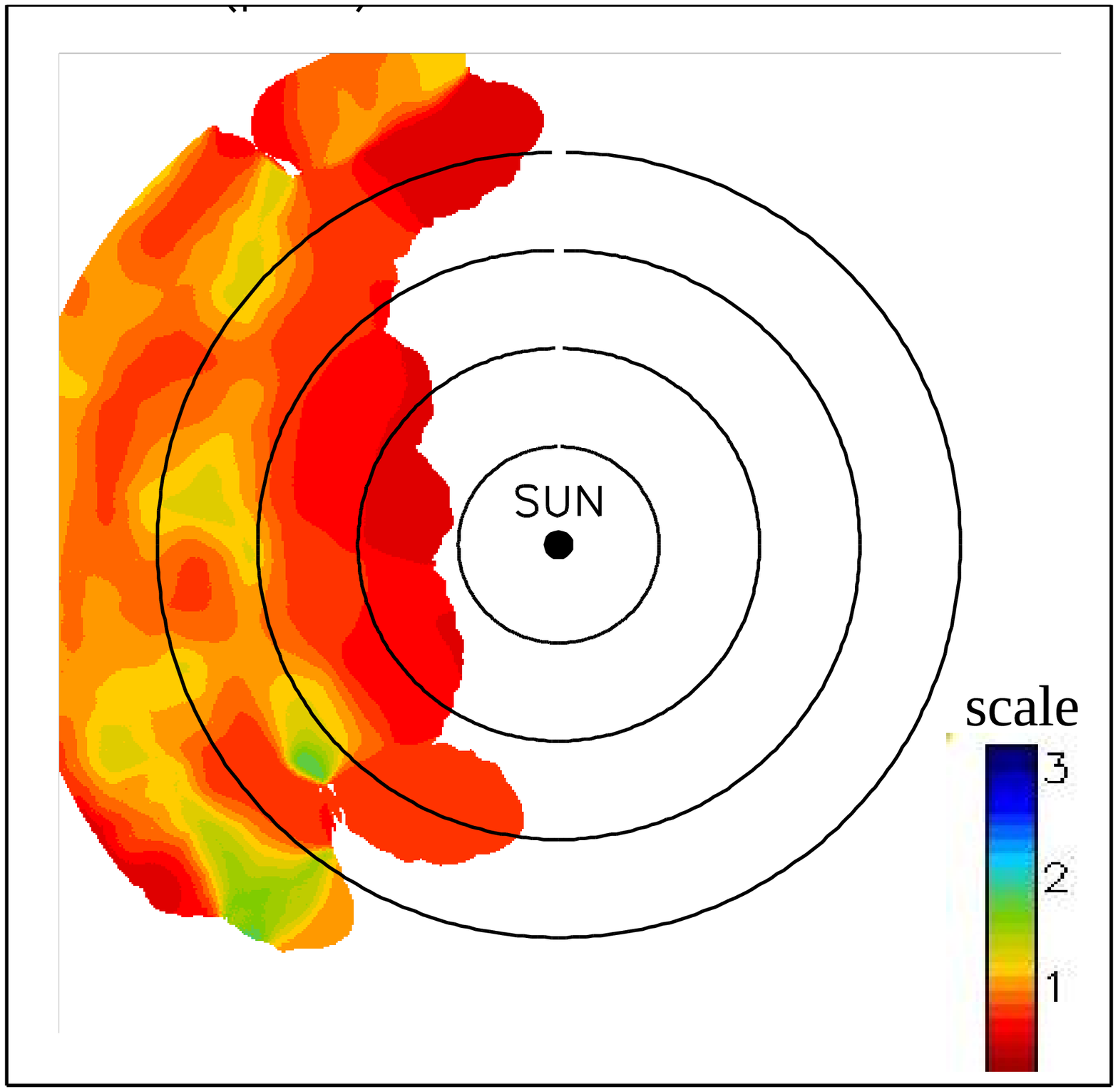,height=4.84cm,width=4.80cm}
\caption{IPS images of disturbance associated with the CME event
on December 13, 2006 and they correspond to IPS measurements
at $\sim$0 UT (left image) and $\sim$14:30 UT (right image) on 
December 14, 2006. In these
`{\it Position Angle} -- {\it Heliocentric Distance}' plots, the 
Sun is at the center.  The concentric circles are 50, 100, 150, 
and 200 {\rsun}. At the LASCO C2 field of view, the CME onset was 
observed at 02:54 UT on December 13, 2006. As seen in the above 
images CME has expanded and moved to larger distances with time.}
\end{figure}

\begin{figure}[t]
\centering
\psfig{file=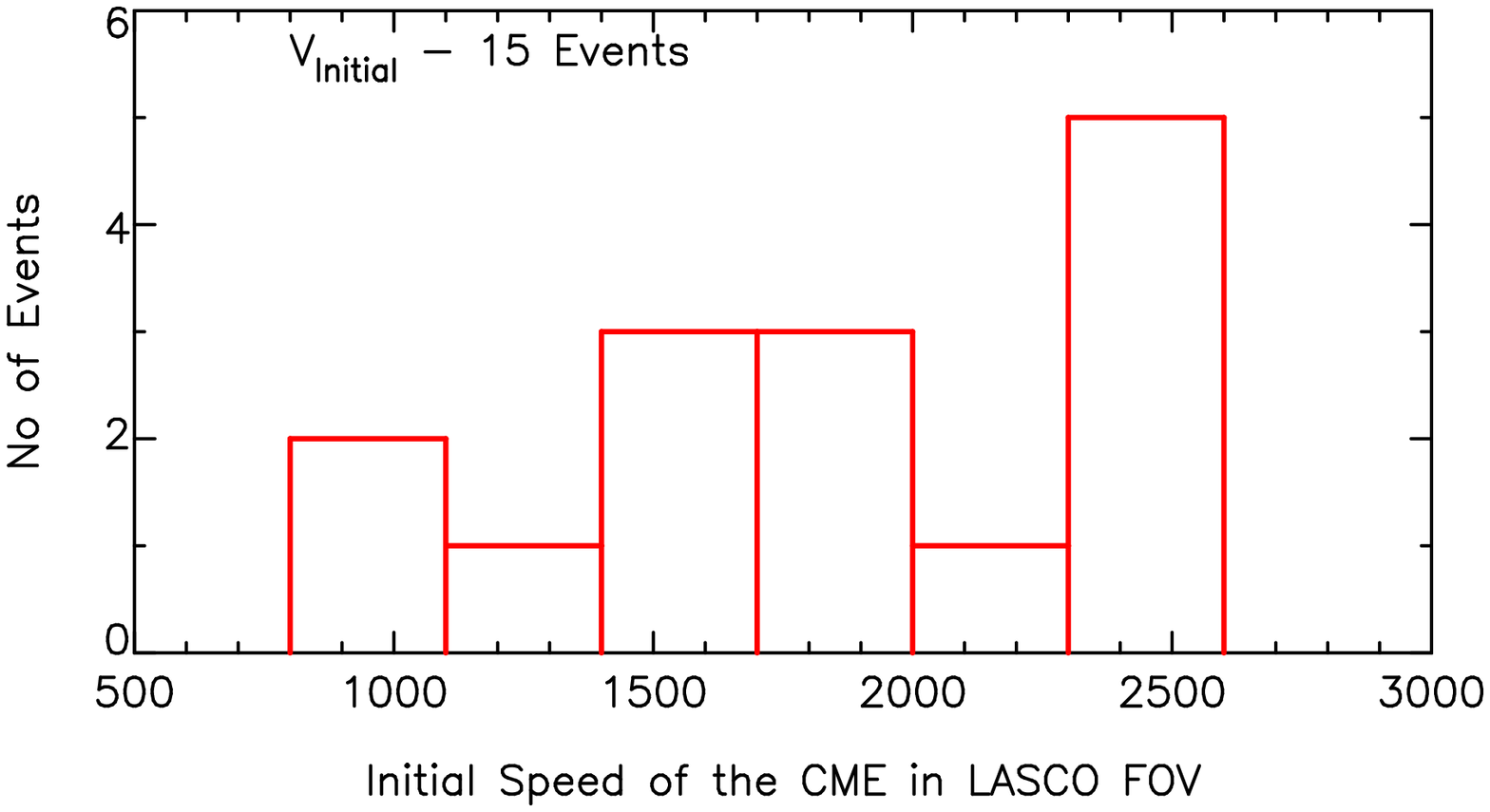,width=5.5cm, angle=-90}\\
\psfig{file=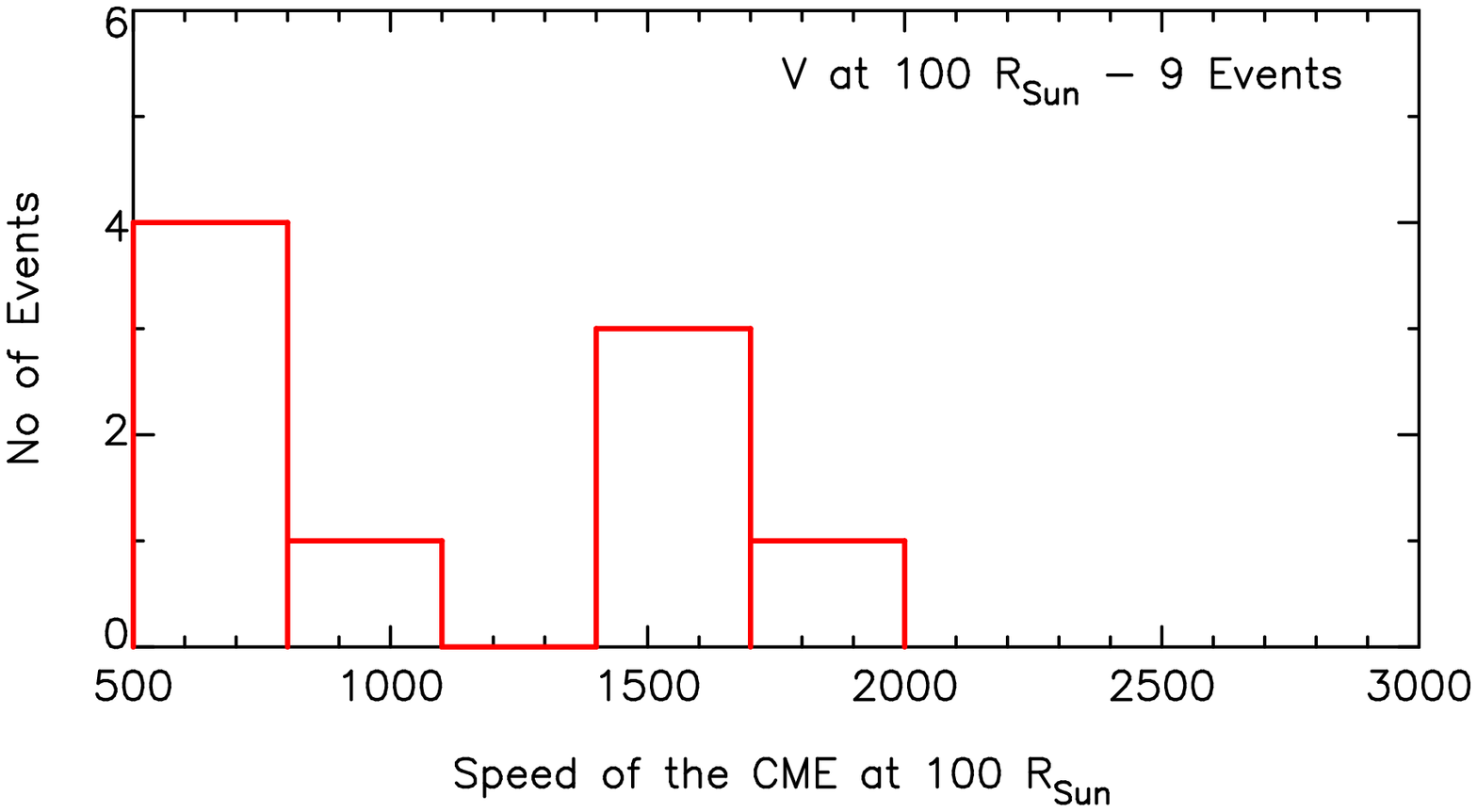,width=5.5cm, angle=-90}
\caption{Histograms of initial speed of the CME in the LASCO
field of view and speed at $\sim$100 {\rsun}. The initial 
speeds are high $\geq$900 {\kmps} and cover a wide range
$\sim$900--2600 {\kmps}. At distance $\sim$100 {\rsun}, the
propagation speed tends to reduce, indicating the energy exchange 
between the CME and solar wind and/or the drag 
force experienced by the CME in the ambient solar wind.}
\end{figure}

\subsection{Interplanetary Scintillation Images}

The interplanetary scintillation (IPS) measurements on a large 
number 
of radio sources are useful to image disturbances associated with 
a propagating CME in the distance range of $\sim$40--250 {\rsun}
\cite{mano01,mano06}{}. Figure 7 shows
sample images of the inner heliosphere associated with the CME
event on December 13, 2006. These are `{\it position angle} --
{\it heliocentric distance}' images and the Sun is at the center. 
The concentric circles are, respectively, 50, 100, 150, and 200
{\rsun}. These images are made from the normalised scintillation index
(denoted by {\it g}) measurements obtained from a grid of large number 
of radio sources ($\sim$1000 sources per day)\cite{mano06}{}.
The increase in the value of normalised scintillation index indicates the 
presence of CME associated disturbance. The color-code scale shown in
Figure 7 from red to blue represents the background solar wind 
({\it g}~$\approx$~1) to an increased level of density and/or density 
turbulence ({\it g}~$\approx$~3).
The image at the left of the figure corresponds 
to IPS measurements on December 14, centered around 0 UT and the 
right image represents the position of CME associated disturbance 
$\sim$14:30 UT on the same day. In each image, the time increases from
the right-hand side of the image (west of the Sun) to the left-hand side 
of the image (east of the Sun). For example, the snapshot of a halo CME at 
a given time will appear as a ring around the Sun \cite{mano06}{}. A halo 
CME observed at the west expands to larger distance with time. The difference
in distance seen between 0 UT and 14:30 UT images shows the propagation of 
the CME. The halo CME expansion/propagation to farther distance with time is 
evident in these images. 

\begin{figure}[t]
\centering
\psfig{file=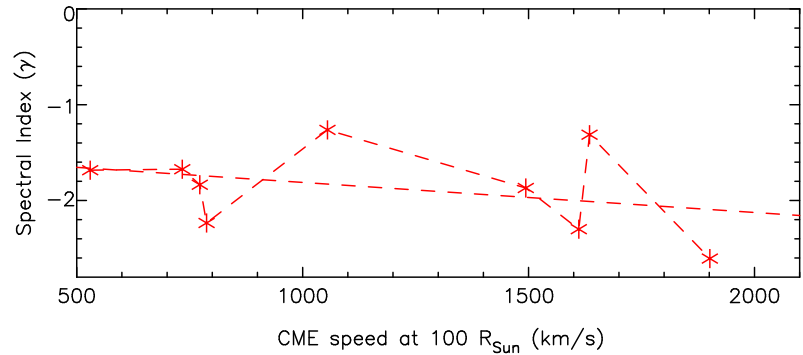,height=10.5cm, angle=-90}
\caption{Spectral index measured at 15 hours after the flare event 
is compared with the IPS speed of the CME at $\sim$100 {\rsun}.}
\end{figure}

As in the case of LASCO images, the above such IPS images can also 
be used to obtain the onset time of CME at different distances from the 
Sun and the `{\it heliocentric distance} -- {\it time}' plot can  
provide the speed profile of the CME in the inner heliosphere\cite{mano06}{}. 
In the present study, we compare the properties of the particle 
events at the near-Sun region (based on LASCO field of view speed) and
at $\sim$100 {\rsun} (based on IPS estimates). For some selective
events, which have both LASCO and IPS measurements, the histograms of 
initial speed and speed at $\sim$100 {\rsun} are plotted in Figure 8.
The initial-speed plot covers a range 900--2600 {\kmps}. However,
as the CME propagates to $\sim$100 {\rsun}, the speed tends to
reduce. As shown by Manoharan \cite{mano06}{}, the drag
force experienced by the CME due to the interaction with ambient solar
wind essentially slows the CME. In other words, the energy exchange
between the CME and solar wind plays a major role in shaping the
speed profile of the CME with time/distance from the Sun. Moreover, 
it is likely that the energy exchange between the CME and the 
ambient solar wind can be quick for an extremely high-speed event
\cite{mano06}{}.

In Figure 9, the spectral index observed at 15 hours after the solar 
event has been compared with speed obtained from IPS at $\sim$100 {\rsun}. 
Although this plot includes only 9 events, it shows similar trend seen 
in the near-Sun region (refer to Figure 6). The range of speeds at this
distance tends to narrow to the lower side and the energy spectrum is
steeper than that of the near-Sun spectrum.

\section{Summary and Conclusion}

In this study, we have selected 31 intense SEP events associated with 
GLEs occurred during 1989 -- 2006. All these gradual particle events 
had association with intense flares/CMEs, which originated towards the 
western side of the Sun (at longitude $>$10$^\circ$ east), confirming
the importance of the magnetic connectivity between the particle 
acceleration source region and the observing point at the Earth. These 
CMEs are fast and  wide (halos and partial halos having width $>$180$^\circ$).
 Their speeds cover the range between 900 and 2600 
{\kmps}. The associated flares are also intense, most of them having
X classification. The results show that the energy spectrum in the 
range 10--100 MeV is flat near the peak of the particle flux profile 
and as the CME propagated outward from the Sun, the spectrum steepens. 

As the CME propagates, the high-energy particle flux decreases quickly 
than the low-energy particle flux, which shows that at the shock front 
the acceleration of low-energy particles is efficient. The rate of 
steepening of the spectrum varies from one event to the other, 
suggesting the energy available with each event is determined by the 
internal energy and the CME-driven shock to accelerate the particle.

The spectral index compared with the initial speed of the CME in
the LASCO field of view shows rather marginal steepening with speed.
The similar trend is observed at 6 and 15 hours after the flare onset.
Also in the IPS field of view, at $\sim$100 {\rsun}, the spectrum
is steeper than the 6-hour spectrum. Further the speed 
distribution of CMEs at the LASCO field of view (R $\le$ 30 {\rsun}) 
in the range 900--2600 {\kmps} tends to narrow to $<$2000 {\kmps}
at $\sim$100 {\rsun}. Most of the events attain speeds below 750 {\kmps}. 
It essentially shows the effect of drag force applied by the ambient 
or background solar wind in slowing down the CME-driven shock. However
further investigations are required to understand the dependence of particle
spectrum on the initial speed of the CME.

%\vskip 0.1in 
%\small{
\section*{Acknowledgments}

The authors acknowledge the observing/engineering staff of 
the Radio Astronomy Centre for help in making the IPS observations.
We are grateful to the GOES team for making the particle data available 
on the website. SOHO is a project of international cooperation between 
ESA and NASA. The authors would like to thank for the excellent LASCO-CME 
Catalog generated and maintained by the
Center for Solar Physics and Space Weather, Catholic University of America, in 
cooperation with NASA and Naval Research Laboratory. This work is partially 
supported by the CAWSES-India Phase II Program, sponsored by the Indian Space 
Research Organisation (ISRO).
%}

\end{document}